\documentclass[useAMS,usenatbib,usegraphicx]{mn2e}

\usepackage{amsmath}
\usepackage{url}

\title[Dark matter vs. pulsars]{Dark
matter vs. pulsars: Catching the impostor}
\author[N. Mirabal]{N. Mirabal$^{1,2}$\thanks{E-mail:
mirabal@gae.ucm.es}\\
$^{1}$Ram\'on y Cajal Fellow\\
$^{2}$ Dpto. de F\'isica At\'omica,
Molecular y Nuclear, Universidad Complutense de
Madrid, Spain\\
}

\begin{document}

\date{}

\pagerange{\pageref{firstpage}--\pageref{lastpage}} \pubyear{2013}

\maketitle

\label{firstpage}

\begin{abstract}
Evidence of excess GeV emission nearly coinciding with the 
Galactic Centre
has been  interpreted as a possible signature of
annihilating dark matter. 
In this paper, we argue that it seems too early  
to discard pulsars as a viable explanation
for the observed excess. On the heels of the recently
released Second {\it Fermi} LAT
Pulsar Catalogue (2FPC), it is still 
possible that a population of hard ($\Gamma < 1$) millisecond pulsars (MSPs)
either endemic to the innermost region
or part of a larger nascent collection of hard MSPs that appears to be
emerging in the 2FPC could 
explain the GeV excess near the Galactic Centre.
\end{abstract}


\section{Introduction}
At first glance, pulsars and dark matter appear to have nothing in common, 
the former 
are magnificent spinning 
neutron stars with impeccable timing \citep{bell,gold}, while 
the latter embodies
the most profound mystery at the crossroads
of gravity and particle physics \citep{peebles}. 
But, on closer inspection, one actually 
realises that they share more than meets the
eye. \citet{baltz} recognised this seemingly innocuous conflict
when they noted that pulsars would be one of
the biggest obstacle to proving a dark mater astrophysical 
signal. An avalanche of recent results has just 
reinforced the ambiguity
\citep{aharonian,cholis}.

This would be purely anecdotal were it not for the fact that
we have not identified a 
dark matter culprit.  The underlying reason is 
that pulsars and dark matter are predicted to share
similar spectral signatures with sharp cutoffs, 
despite dramatically different astrophysical 
origins. Around pulsars, gamma-ray photons are emitted via 
curvature radiation of accelerated particles with an exponential cutoff
at the maximum curvature 
energy around a few GeV \citep{rybicki,2p}.  
In contrast, a number of dark matter models predict that cosmic dark  
particles
will annihilate into known elementary particles that will subsequently  
generate secondary photons. The resulting gamma-ray spectrum should show a
cutoff near the dark matter particle mass $m_{\chi}$ 
\citep{berg,bring}. 

This issue has come to bear on current searches for dark matter in the  
purlieus of the Galactic Centre. The central concentration of dark matter is
arguably the most promising place to search for unusual 
annihilation products. As it turns out, over the past few years
a number of groups have
noticed the presence of excess GeV 
emission around the Galactic Centre \citep{hl,boyarsky,
abazajian}.
Whilst these results are possible breakthroughs in dark
matter research, the region over which the excess GeV emission has been found 
is scientifically daunting with local sources of  
diffuse emission and unresolved gamma-ray emitters that can
easily sequester any secondary emission associated with dark matter.

Procedurally, 
a final confirmation of dark matter annihilation must exclude all other
available astrophysical explanations.  
Exploiting 
the spectral shape  of the excess GeV emission, \citet{abazajian} and 
\citet{hl} have concluded 
that known gamma-ray pulsars cannot account for such a  signal.
Most arguments against pulsars have been built around the
premise that the spectral shape of {\it Fermi} pulsars ($0.4 < \Gamma < 2.0$)
cannot account for the much harder ($\Gamma \approx 0.5$) spectrum
of the GeV
excess \citep{hl}.
These studies lead to the 
seemingly unavoidable conclusion that we are detecting dark matter 
annihilation.
The true situation is more complicated.

In their favour, the exponential cutoff in pulsars has been 
measured exquisitely well \citep{2p}. 
The observed peak for the excess GeV emission
at 1--4 GeV is consistent with
the observed cutoff energy for {\it Fermi} pulsars 
that tend to cluster around 0.4 GeV $< E_{\rm cutoff} <$ 6 GeV. 
The average {\it Fermi} gamma-ray luminosity for  
MSPs from six months of {\it Fermi} data 
also appears to be in the right ballpark of the excess 
\citep{wharton}. Unlike 
dark matter, the existence of a Galactic Centre population of  
neutron stars has also been firmly established
based on discovery of a handful of pulsars within $15^{\prime}$ of
Sgr A$^{*}$ \citep{muno,deneva}. 

Thus, although there are some differences at the astrophysical level,
observationally it is still very difficult 
to tell pulsars and dark matter apart for 
models with dark matter particle mass $m_{\chi}$ between
0.1 and 100 GeV. Here we suggest that a  
population of hard ($\Gamma < 1$) MSPs could still account for 
the GeV excess at 
the Galactic Centre.
Initially, we estimate the pulsar population needed within 
a few degrees of the Galactic Centre in the context 
of the newly released second {\it Fermi} Large Area Telescope (LAT) 
pulsar catalogue \citep{2p}. 
Next, we motivate and discuss potential reasons 
for a concentration of hard MSPs in the innermost region.
We close with implications and possible ways forward.

\section{The Galactic Centre by the numbers}
\label{smallviewing}
\begin{figure}
\hfil
\includegraphics[width=3.5in]{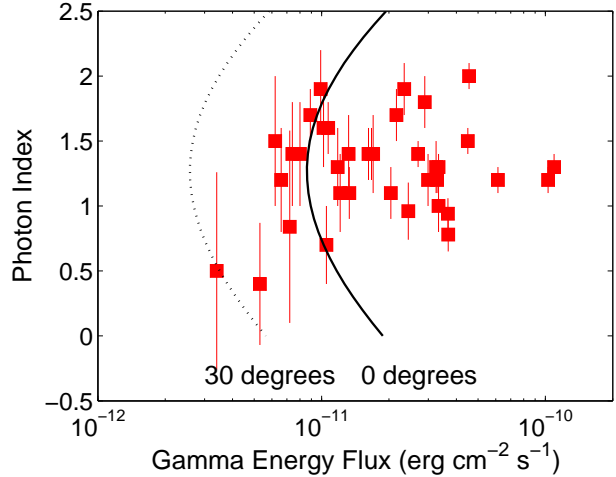}
\hfil
\caption{Power law index versus gamma-ray
energy flux $G_{100}$ for MSPs. The solid line
denotes the effective sensitivity curve for pulsar spectra with an
exponential cutoff energy of $E_{\rm cutoff} = 2$ GeV at $b = 0^{\circ}$.
 The dotted curve is for high-latitude pulsars at  $|b| = 30^{\circ}$.}
\label{figure1}
\end{figure}

Our first task is to revise the measured pulsar luminosities and 
certify that it is possible to reproduce the 
excess GeV emission at
the Galactic Centre with the most recent list of 
{\it Fermi} MSPs.  
Using three years of data, the Second {\it Fermi} LAT  
Pulsar Catalogue (2FPC) reports a total of 
117 gamma-ray pulsars of which 77 are 
young or middle-aged and 40 are MSPs \citep{2p}. 
The updated catalogue nearly triples the previous 
{\it Fermi} pulsar list and it appears to be
progressively populated by more MSPs with harder 
photon index $\Gamma < 1$. This is illustrated in
Figure \ref{figure1}, where we plot the 
photon spectral index against the 
energy flux from 0.1 to 100 GeV for {\it Fermi} MSPs.

As in \citet{hl}, we adopt a 0.1 -- 100 GeV energy flux of
$f_{\rm GC} \approx 8 \times 10^{-10}$ erg cm$^{-2}$ s$^{-1}$ 
for the GeV excess from the Galactic Centre. 
At the distance of the Galactic Centre (8.3 kpc), 
this corresponds
to a gamma-ray luminosity $L_{\rm GC} \approx 6.6 \times 10^{36}$ erg
s$^{-1}$ $f_{\Omega}$, 
where $f_{\Omega}$ represents 
the correction factor. Following \citet{wharton}, we can estimate
the total number of MSPs potentially present in the Galactic Centre 
from the gamma-ray luminosity $L_{\rm GC}$ and the average
luminosity of a typical pulsar $L_{\gamma}$ using,

\begin{equation}
N_{\rm MSP} = \frac{L_{\rm GC}}{L_{\gamma}}.
\end{equation}

In order to solve this relation, we must first understand the
current LAT sensitivity to pulsars. 
Starting with the predicted sensitivity limits for a pulsar-like spectrum
with $\Gamma= 1.8$ and $E_{\rm cutoff} = 2$ GeV
reported by \citet{2p}, 
we have built a sensitivity curve for 
the entire range of power law indices assuming a pulsar-like
exponential cutoff energy spectrum, with a fixed parameter
$E_{\rm cutoff} = 2$ GeV  
\begin{equation}\label{expcutoff}
G_{100} = \int_{\rm 0.1 \, GeV}^{\rm 100 \, GeV} K E^{1-\Gamma} \exp \left(\- \frac{E}{E_{\rm cutoff}} \right) \
{\rm d} E.
\end{equation}

Figure \ref{figure1} shows the LAT sensitivity from 0.1 to 
100 GeV for $b = 0^{\circ}$ and
$|b| = 30^{\circ}$. Clearly, there is a strong
latitude dependence of the sensitivity. In addition, 
the sensitivity  
degrades by a factor of $\approx 2$ for harder pulsar spectra. 
Another source of uncertainty is the diffuse flux  in
the innermost region. Only 5  
out of 40 MSPS are located at 
$|\ell| < 5^{\circ} or~ |b| < 5^{\circ}$, including 3 of the most luminous MSPs 
in the entire sample. 
In short, we have not fully resolved the MSP population in the Galactic
Centre region. 

Assuming a flux threshold of 
$\approx 10^{-11}$ erg cm$^{-2}$ s$^{-1}$ for a photon 
index $\Gamma \approx 0.5$ 
implies that 
$N_{\rm MSP} \approx 80$  
are needed to reproduce the excess GeV emission from the Galactic Centre.
For a more conservative estimate, 
we can take PSR~J1600--3053  as a typical representative of the 
hard-spectrum pulsar population at the Galactic Centre. With a  
luminosity $L_{\gamma}$ = $1.7 \times 10^{33}$ erg s$^{-1}$ and 
a photon index $\Gamma = 0.4$, PSR~J1600-3053 
has the hardest pulsar spectrum of the MSPs 
in the 2FPC \citep{2p}. A population of 
comparable luminosity 
would raise the requirement to $N_{\rm MSP} \approx 3900$. 
Both estimates are still in agreement with the population  
of predicted MSPs ($few\times 10^{3}$) at the Galactic Centre derived
from observations at other wavelengths 
\citep{deneva,wharton}. 

This does not necessarily 
imply that all MSPs in the Galactic Centre region
will be gamma-ray emitters. Of the 169 known 
field radio MSPs, about 40 have been detected by {\it Fermi} \citep{2p}.
Therefore at least 20\% of MSPs should be detectable in gamma rays.
But at this point, 
it seems that we have only scratched the tip of iceberg in terms of
MSP detections and could be missing the bulk of these systems \citep{story}.

\section{Nurture or pulsar oddities?}
In terms of luminosity, MSPs appear to be viable explanation for
the inner gamma-ray excess. However, we need to further
motivate the presence of a population of
hard-spectrum MSPs at the Galactic Centre.  
From a theoretical standpoint,
spectral variations are expected in the photon index
depending on the viewing geometry
and the contribution from different emission regions \citep{hirotani,takata2}.
In contrast, annihilating dark matter should be spectrally
invariant across the sky.

Our analysis thus far admits two possible MSP scenarios. The first is
that  {\it Fermi} is detecting an MSP
population that is truly unique to the Galactic Centre. 
Based on EGRET observations, 
\citet{wang} argued that most MSP pulsars near the
Galactic Centre are formed from old, slow moving neutron stars 
that have been recycled
to MSPs. Because of its high stellar density ($10^3$--$10^6$ stars pc$^{-3}$), 
only dense globular clusters ($> 10^3$ stars pc$^{-3}$) with a long
dynamical history can come close to mimicking
the Galactic Centre neighbourhood. Steady gamma-ray emission has been
significantly detected towards a growing population 
globular clusters \citep{abdo}. These tend to show hard
spectral indices ($0.7 < \Gamma < 1.7$) and exponential cutoffs in
the range 1.0--2.6 GeV, which go in the right direction
to explain the excess GeV emission. 
 
\begin{figure}
\hfil
\includegraphics[width=3.5in]{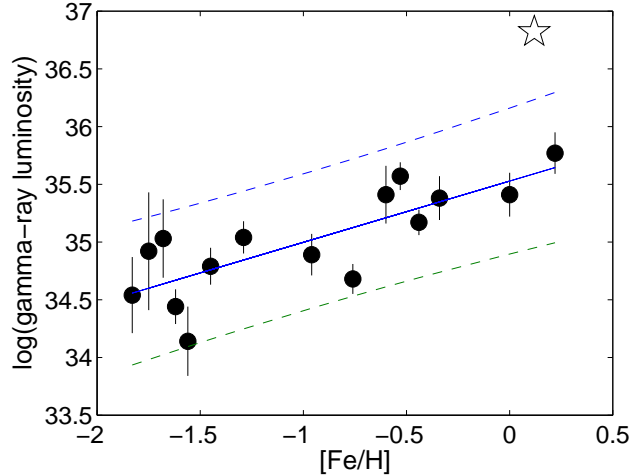}
\hfil
\caption{Observed gamma-ray luminosity for {\it Fermi} globular
cluster vs. metallicity [Fe/H]. 
The straight line shows the fundamental plane relationship
from \citet{hui}.
The dashed lines represent 95\% confidence bands.
The star shows the Galactic Centre, consistent with the
tendency for metal-rich environments and higher gamma-ray
luminosity but clearly an outlier from the reported relationship.}
\label{figure2}
\end{figure}

Given the similarities in stellar densities, 
we want to test whether the Galactic
Centre is consistent with the properties of 
gamma-ray-emitting globular clusters.
One of the observational properties of globular clusters 
that might provide context is the apparent trend for
higher gamma-ray luminosity with increasing [Fe/H] for globular clusters
\citep{hui}. 
Figure \ref{figure2}
marks the location where the Galactic Centre falls 
with respect to the fundamental
plane of globular clusters derived by \citet{hui}. From the figure, 
we see that the   
Galactic Centre appears to be incompatible with the fundamental-plane
relationship.  Apart from the particular stellar dynamics around the 
supermassive black hole in the Galactic Centre, 
the difference 
may be  due to a near solar metallicity of the Galactic Centre
[Fe/H] = 0.12 \citep{ramirez} and the presence of compact young 
clusters in the central 50 pc that can reach central densities as high as 
$10^6$ stars pc$^{-3}$ \citep{figer}.
This could be the first tentative indication that we are 
dealing with an endemic MSP population.

As yet, the formation channels for MSPs remain a puzzle.
It is generally agreed that close, interacting 
X-ray binaries eventually end up as MSPs \citep{bhatta}.  
Using population synthesis models, 
\citet{belczynski} argued that  Roche lobe overflow (RLOF) systems 
involving the collapse of massive ONeMg white dwarfs should be
pervasive in the Galactic Centre. Metal-rich stars
fill their Roche lobe more easily 
\citep{ivanova}, and 
as a result the formation rate of MSPs could be enhanced compared
to globular clusters. This would explain the deviation of
the Galactic Centre from
the fundamental plane of globular clusters. 

As for the hard spectrum measured by \citet{hl}, 
we note that the median photon index 
of the X-ray sources discovered by {\it Chandra}
within the inner $9^{\prime}$ of the Galaxy
is $\Gamma = 0.7$ \citep{muno}.  
In the 2FPC, one sees a growing
trend $\Gamma \approx \dot{E}^{0.4}$ for harder spectrum
at lower MSP spindown luminosity  \citep{2p}.
The spindown luminosity can be written 

\begin{equation}
\dot{E} \propto (\mu^2 \Omega_*^4/c^3) (1+\sin^2\alpha),
\end{equation}

\noindent
where $\mu$ is the dipole moment, $\Omega_*$ is the
rotation frequency, 
and $\alpha$ is the magnetic inclination angle \citep{spitkovsky}.
One simple prescription is that
the MSP formation process near the Galactic Centre favours
smaller magnetic inclination angles $\alpha \approx 0$.
As argued by \citet{johnson}, this might be a natural
tendency for recycled gamma-ray pulsars in general, but
could be more frequent in the Galactic Centre. 
If recycled pulsars with 
ONeMg companions are prevalent in the Galactic Centre region 
\citep{belczynski}, they will be 
much slower rotators (lower $\dot{E}$) than
MSPs with He WD companions \citep{tauris}.

Since there appears to be an emerging population of hard spectra 
outside the Galactic plane, the alternative scenario is that hard MSPs 
are not necessarily tied to the Galactic Centre, but have formed
throughout the Galaxy. 
If true, the incipient hard ($\Gamma < 1$) sample
might reveal some clues about their origin. Looking at the seven 
hard {\it Fermi} pulsars individually, there is a smorgasbord of 
pulsar oddities. Six of these seven MSPs are in binary systems. 
PSR J1614--2230 hosts the most massive pulsar known to date \citep{demorest}.  
PSR J2051--0827 has one of the shortest orbital periods 
$P_{b} \approx 2.4$ hr \citep{stappers},
while PSR J2302+4442 has one of the longest
$P_{b} \approx 125.9$ days \citep{cognard}. PSR J1600--3053 is
among the best high-precision pulsars known \citep{ord}. PSR J2124--3358 is 
the lone isolated MSP \citep{mignani}. PSR J0101--6422 is the only
object where simple geometric emission models fail to explain the
observed peaks, suggesting that the details of its MSP magnetosphere
are more complex than expected \citep{kerr}. Because of frequent
encounters and companion exchanges, the Galactic Centre
could be more conducive to the production of pulsar oddities.
When compared with other field MSPs, 
no single connecting thread stands out in this bunch. 
However, we cannot rule out that these systems 
were formed in rare special 
environments. Dedicated studies of possible MSP
birth locations and
companions could reveal additional information.

\section{Implications and conclusions}
\label{interp}
In view of our results, it is still possible
to explain the gamma-ray excess in the Galactic Centre with a 
population of hard-spectra MSPs.
An essential test of these ideas is to search for similar excesses in other
sections of the Galaxy. Interestingly, a 
possible excess coincident with the {\it Fermi} bubbles has been reported 
\citep{hs,urbano}. 
Also \citet{ackermann} indicated that current diffuse
gamma-ray models under predict the data in the Galactic
plane. It
is possible that some of the MSPs discussed here might have migrated to 
regions adjacent to the Galactic Centre. 
Even if MSPs are not the culprits of the excess, 
inverse Compton scattering of diffuse X-ray emission 
near the Galactic Centre \citep{muno2} 
by relativistic electrons (Lorentz factors $\gamma \approx$ 500--1000)
could leave an imprint at GeV energies.
A deep Galactic Centre survey with the ability to resolve hard sources 
planned for the Cherenkov Telescope Array should help clarify this issue 
\citep{cta,dubus}.

It seems clear that modelling the unresolved 
pulsar distribution will be critical step
to assess astrophysical signatures of dark matter. 
Beyond the Galactic Centre, pulsars above the Galactic plane
could potentially
mimic Galactic dark matter subhalos \citep{baltz,mirabal}. Nearby pulsars
might also provide a source for the observed rising positron fraction
\citep{grasso,hoop}. As a result, rather than treating excess GeV emission 
as evidence for dark matter, it now seems obligatory to start including 
undetected pulsars as one of the largest contributors to this 
complex signal \citep{gordon}. This has been done rather successfully
with the diffuse gamma-ray emission from the 
interstellar medium  \citep{ackermann}.
An excellent first attempt to account for pulsars 
was advanced by \citet{geringer}. 
Since it is not obvious how to include a population that 
might be generally undetectable to {\it Fermi} surveys, 
we must be extremely meticulous in building a composite
pulsar template \citep{cholis2}. 
Perhaps it will be found that pulsars and dark matter contribute to
the GeV excess. With improved techniques, we could start
disentangling the dark signal.

From the data reported so far, it is tempting to
conclude that we are starting to see the first signals
of annihilating dark matter. But we must take this road 
with caution in view of the degeneracy with pulsars.
Technically, a population of MSPs with steep spectrum 
would be very difficult to probe in 
radio \citep{wharton}. 
The dearth of photons 
near the LAT sensitivity would also make it very difficult to
conduct gamma-ray blind period searches with {\it Fermi} \citep{pablo}. 
None the less, a dedicated search 
for gamma-ray pulsations from the inner Galaxy is
a must \citep{saz}. A
radio survey for additional pulsars with the next generation of
sensitive receivers also
appears to be a necessity.

At the end of the day, the strongest astrophysical 
case for dark matter annihilation 
will be able to convene multiple sources with the same spectral signature 
across the sky. A tie-break would be the localisation 
of ``Crab-like'' power-law tails in {\it Fermi} pulsars at energies above
20 GeV with the upcoming Cherenkov Telescope Array
\citep{hassan,emma}. An alternative possibility is the direct 
detection of a spatially extended dark matter source  \citep{bring}. 
A line-like gamma-ray feature 
would be a game point \citep{torsten,wen,su1,fink,hektor,acker13}.
Confirmation in at least two pillars of dark matter detection
clinches the game \citep{bauer}. 
However, indirect dark matter detection may prove much more subtle. 
Observing the Galactic Centre more frequently as part of 
a renewed {\it Fermi} observing strategy 
could start to break the stalemate \citep{weniger}.

\section*{Acknowledgments}
We thank K. S. Cheng for referring us to the \citet{wang} paper. 
We acknowledge Dan Hooper and Per Olof Lindblad for helpful email exchanges. 
We also thank the referee for useful suggestions and comments on the 
manuscript.
N.M. acknowledges support from the Spanish taxpayers 
through a Ram\'on y Cajal fellowship and the 
Consolider-Ingenio 2010 Programme under grant MultiDark CSD2009-00064.

\label{lastpage}
\end{document}